\documentstyle[12pt]{article}
\input{epsfig} \textheight 21 cm \textwidth 16 cm \oddsidemargin
-0.15 cm
\begin{document}


\title{Resonances in one-dimensional Disordered Chain }

\author{Herv\'{e} Kunz$^1$ and Boris Shapiro$^2$\\
$^1$Ecole Polytechnique F\'{e}d\'{e}rale de Lausanne,\\ ITP, PHB
Ecublens, CH-1015 Lausanne, Switzerland\\
$^2$Department of Physics, Technion-Israel Institute of
Technology, Haifa 32000, Israel}
\maketitle

\begin{abstract}
We study the average density of resonances, $\langle
\rho(x,y)\rangle$, in a semi-infinite disordered chain coupled to
a perfect lead. The function $\langle \rho(x,y)\rangle$ is defined
in the complex energy plane and the distance $y$ from the real
axes determines the resonance width.  We concentrate on strong
disorder and derive the asymptotic behavior of $\langle
\rho(x,y)\rangle$ in the limit of small $y$.
\end{abstract}

\vskip 4cm
\section{Introduction }
Open quantum systems often exhibit the phenomenon of resonances.
Resonances correspond to quasi-stationary states which have a long
life-time but eventually decay into the continuum. (A particle,
initially within the system, escapes to infinity.) They are
characterized by complex energies, $\tilde E_\alpha = E_\alpha
-{i\over 2}\Gamma_\alpha$, which correspond to poles of the
S-matrix on the unphysical sheet of the complex energy plane
\cite{LL,Zeld}. There are many examples of resonances in atomic
and nuclear physics. More recently, there has been much interest
in resonant phenomena in the field of chaotic and disordered
systems (for recent reviews see \cite{ Mir, Kot1}).

There is  considerable amount of work concerning the distribution
$P(\Gamma)$ of  resonance widths in one-dimensional disordered
chains \cite{TEG, Comt, Orl, Tit, Kot2}. Numerical studies
presented in that work demonstrate that, in a broad range of
$\Gamma$, $P(\Gamma)\propto \Gamma^{-\gamma}$, with the exponent
$\gamma$ being close to 1. (This behavior is not restricted to
disordered chains, but pertains also to two- and three-dimensional
systems with { \em localized} states \cite{Orl, Kot2}). An
intuitive argument, which assumes a uniform distribution for the
localization centers of exponentially localized states, leads to a
$(1/\Gamma)$-behavior \cite{Orl, Tit, Kot2}. The analytical
calculation in \cite{Tit}, performed for a one-dimensional
continuous (white-noise) potential, exhibits this behavior for
sufficiently small $\Gamma$, followed by a sharp cutoff at still
smaller $\Gamma$, due to the finite size of the sample. It is also
shown in \cite{Tit} that in a broad range of $\Gamma$, $P(\Gamma)$
is well fitted by a function $\Gamma^{-1.25}$. In the present
Letter we develop an analytical approach for a  discrete,
tight-binding random chain. We treat the problem in the limit of
{\em strong} disorder and derive the asymptotically exact
$(1/\Gamma)$-behavior for a {\em semi-infinite} system.

\section{The model and its Effective Hamiltonian}
We consider a semi-infinite disordered chain coupled to a
(semi-infinite) perfect lead.  $n=1,2,....$, designate sites along
the  chain. Each site of the chain is assigned a site energy,
$\epsilon_n$. Different $\epsilon_n$'s ($n=1,2,....$) are
independent random variables chosen from some symmetric
distribution $q(\epsilon)$. Sites of the lead are labelled by
$n=0,-1,-2,....$. All sites of the lead are assigned $\epsilon_n =
0$. The lead simulates the free space outside the chain. All
nearest neighbor sites of the full system (chain + lead) are
coupled to each other by a hopping amplitude $t$, so that a
particle, initially located somewhere within the chain, will
eventually escape into the lead.

 The most direct approach to the problem of resonances amounts to solving the stationary
Schr\"odinger equation, for the entire system,  with the boundary
condition of an {\em outgoing wave} only.
This
condition, which makes the problem non-Hermitian, describes a
particle ejected from the system. The Schr\"odinger equation with
such boundary condition admits complex eigenvalues $\tilde
E_\alpha$, which correspond to the resonances \cite{LL,Zeld}. This
kind of approach, which leads in a natural way to a non-Hermitian
effective Hamiltonian, has been used for a long time in scattering
theory, including scattering in chaotic and disordered systems
(\cite{ Kot1, Datta}, and references therein). For our system
the approach amounts to solving the infinite set of
coupled equations
\begin{equation}\label{schrod1d}
-t\psi_{n+1} - t\psi_{n-1} + \epsilon_n\psi_n = \tilde E
\psi_n\quad\quad (-\infty <n<\infty)
\end{equation}
We recall that $\epsilon_n=0$ for $n<1$ (the lead) and it is
 random for $n\geq 1$ (the chain).
 Eqs.(\ref{schrod1d}) are to be solved subjected to the boundary
condition $\psi_n\propto \exp -i\tilde k n$ , corresponding to an
outgoing wave in the lead. The complex wave vector $\tilde k$
determines $\tilde E$ according to $\tilde E = -2t \cos \tilde k$.
It is straightforward to eliminate from Eqs.(\ref{schrod1d}) all
$\psi_n$'s with $n<1$ , thus reducing the problem to a system of
equations for the amplitudes $\psi_n$ on the sites of the
disordered chain alone ($n=1,2,....$).
\begin{equation}\label{nhschrod1d}
-t\psi_{n+1} - t\psi_{n-1} + \tilde\epsilon_n\psi_n = \tilde E
\psi_n\quad\quad (n = 1,2,\ldots  )\,
\end{equation}
with the condition $\psi_0=0$. Here $\tilde\epsilon_n =
\epsilon_n$ for $n=2,3, ....$, but not for $n=1$. This end site is
assigned  a complex energy $\tilde\epsilon_1 = \epsilon_1 -t\exp
i\tilde k$ which describes coupling to the outside world.
 Thus, the
resonances are given by the complex eigenvalues of the
non-Hermitian effective Hamiltonian defined in (\ref{nhschrod1d}).
Note that Eq. (\ref{nhschrod1d}) does not constitute a standard
eigenvalue problem because $\tilde\epsilon_1$ contains $\tilde k$,
which is related to $\tilde E$. The problem is often simplified by
fixing $\tilde k$ at some real value, consistent with the value of
energy near which one is looking for resonances. Such
simplification corresponds to replacing the ``exact resonances" by
``parametric" ones  \cite{TEG}. There are indications that for
sufficiently narrow resonances the ``parametric" and the ``exact"
distributions are close to one another. In this Letter we restrict
ourselves to ``parametric" resonances. For instance, close to the
middle of the energy band we set $\tilde k = \pi/2$, thus arriving
at the effective Hamiltonian
\begin{equation}\label{Heff}
\left(H_{eff}\right)_{nm} = \tilde\epsilon_n\delta_{nm} -
t_{nm}\quad\quad (n = 1,2,\ldots  )\,.
\end{equation}
where $t_{nm} = t$ for nearest neighbors (and zero otherwise), and
$\tilde\epsilon_{1} = \epsilon_{1} -it$. Let us repeat that all
site energies, except for $\tilde\epsilon_{1}$, are real. The
imaginary part $-it$  of $\tilde\epsilon_{1}$ accounts for the
coupling of the chain to the lead, via the hopping amplitude $t$
connecting site 1 to site 0.  It is convenient to slightly
generalize the model by assigning to this particular amplitude  a
value $t'$, which can differ from all the other hopping amplitudes
$t$. This allows us to tune the coupling from $t'=0$ (closed
chain) to $t'=t$ (fully coupled chain). In what follows we set
$t=1$ and denote the dimensionless coupling strength $v\equiv
t'/t$.

\section{The Average Density of Resonances}
Resonances correspond to  complex eigenvalues of $H_{eff}$, i.e.,
to the poles of the resolvent
\begin{equation}\label{resolvent}
\tilde{\bf G} = {1\over z-H_{eff}}
\end{equation}
in the lower half of the complex energy plane. We denote these
poles by $z_\alpha = x_\alpha + i y_\alpha$. (These are just the
complex energies $\tilde E_\alpha = E_\alpha -{i\over
2}\Gamma_\alpha$, of Section 1, in units of $t$ which we set to
1.) The average density of these poles  is defined as
\begin{equation}\label{DOR}
\langle \rho(x,y) \rangle = \langle \sum_\alpha \delta
(x-x_\alpha)\delta (y-y_\alpha)\rangle,
\end{equation}
where $\langle\cdots\rangle$ denotes disorder averaging, i.e.,
averaging over all realizations of the set $\{\epsilon_n\}$ of the
random site energies.

In order to appreciate the difference between $\langle \rho(x,y)
\rangle$ and the probability distribution, $P(x,y)$, of resonance
width $y$ (for some fixed $x$) let us consider for a moment a
finite chain, of $N$ sites. In this case $P(x,y)$, by definition,
is equal to $\langle \rho(x,y)\rangle$ divided by $N$. When $N$
increases, $P(x,y)$ ``runs away" towards smaller and smaller
$y$'s, and in the $ N\rightarrow \infty$ limit it approaches
$\delta(y)$. (Indeed, for a semi-infinite chain an eigenstate will
be localized, with probability 1, at  infinite distance from the
open end and, thus, will be ignorant about the coupling to the
external world.) On the other hand,  $\langle \rho(x,y)\rangle$
does have a well defined $ N\rightarrow \infty$ limit, for any
fixed (non-zero) $y$. To clarify this assertion, we begin with a
closed semi-infinite chain ($v=0$). In this case all states are
localized ($y_\alpha=0$)- some close to the end (site 1)and some
further away. When the chain is opened ($v=1$), localized states
turn into resonances. The point is that for fixed $y$ (and $x$)
the main contribution to $\langle \rho(x,y)\rangle$ will come from
states that (for $v=0$) were localized around some optimal
distance $d(x,y)$ from the open end. Therefore distant pieces of
the chain do not contribute to $\langle \rho(x,y)\rangle$ and a
well defined $ N\rightarrow \infty$ limit exists. The $N$
necessary for achieving this limit will, of course, depend on $y$
(and $x$) but the limit will eventually be achieved for any $y$,
however small (different from zero). Thus, although for any finite
$N$, $P(x,y)$ and $\langle \rho(x,y)\rangle$ differ only by the
normalization factor $N$, it is $\langle \rho(x,y)\rangle$ that
has a meaningful $ N\rightarrow \infty$ limit. $\langle \rho(x,y)
\rangle $ may be expressed in terms of the resolvent $\tilde{\bf
G}$ as \cite{electrostatics, fz1}
\begin{equation}\label{Gausslaw}
\langle \rho(x,y) \rangle = {1\over 2\pi}\left( \partial_x +
i\partial_y\right)\langle {\rm Tr} \tilde{\bf G}(x,y) \rangle\,.
\end{equation}

One can interpret the $z_\alpha$'s as the positions of unit
electric point charges in the plane, which give rise to an
electric field ${\bf E} = E_x\hat{\bf x} + E_y\hat{\bf y}$. In
this picture, one has simply ${\rm Tr} \tilde{\bf G}(x,y) =
E_x-iE_y$, and (\ref{Gausslaw}) is then interpreted as the Poisson
equation for the averaged electric field and charge density.
Equation (\ref{Gausslaw}) holds also for  a closed chain, if
$\tilde{\bf G}(x,y)$ is replaced by ${\bf G}(x,y)$, where the
untilded ${\bf G}$ is the resolvent of the Hermitian Hamiltonian
of the closed chain. In this case all the charges must lie on the
real axis. Therefore, for any $y$ different from zero, one can
rewrite (\ref{Gausslaw}) as
\begin{equation}\label{Gausslaw1}
\langle \rho(x,y) \rangle = {1\over 2\pi}\left( \partial_x +
i\partial_y\right)\langle {\rm Tr} (\tilde{\bf G}-{\bf G})
\rangle\,.
\end{equation}
The advantage of this representation is that, although both ${\rm
Tr} \tilde{\bf G}$ and ${\rm Tr}{\bf G}$ diverge in the $
N\rightarrow \infty$ limit, ${\rm Tr} (\tilde{\bf G}-{\bf G})$
remains finite. (Of course, it is still perfectly all right to use
(\ref{Gausslaw}), with the proviso that the derivative with
respect to $y$ is taken before taking the $N\rightarrow \infty$
limit.)

 The electrostatic analogy just mentioned makes it clear that
$\langle{\rm Tr} \tilde{\bf G}(x,y) \rangle$ cannot be a complex
analytic function of $z=x+iy$, but must depend separately on $x$
and $y$. This lack of analyticity makes the computation of
$\langle \rho(x,y)\rangle$ a more difficult  task than that of the
density of states for a closed (Hermitian) system. In the latter
case there is a well known representation of the resolvent
\begin{equation}\label{selfenergy1}
{\bf G_{nn}}(z) =  {1\over z- \epsilon_n - \Sigma_n(z)}\,.
\end{equation}
where $\Sigma_n(z)$ is the self-energy at site $n$. Let us stress
that (\ref{selfenergy1}) applies to a specific realization of
random site energies  $\{\epsilon_n\}$, i.e.,    $\Sigma_n(z)$ is
a random quantity which depends on the set $\{\epsilon_n\}$. The
statistical treatment of $\Sigma_n(z)$ forms the basis of many
studies of the localization problem \cite{LGP}, starting with the
original work of Anderson \cite{PW}. The notion of self-energy can
be generalized to our non-Hermitian problem, and the analog of
(\ref{selfenergy1}) is
\begin{equation}\label{selfenergy2}
 {\tilde{\bf G}}_{nn}(z) = {1\over z- \epsilon_n -
\tilde\Sigma_n(z)}\,.
\end{equation}
This equation is essentially a definition of $\tilde\Sigma_n(z)$.
It is convenient to assign the imaginary term $-iv$ to
$\tilde\Sigma_1(z)$ rather than to the ``bare" energy of the first
site, so that all site energies, $\{\epsilon_n\}$, in (\ref
{selfenergy2}) are real and are the same as for the corresponding
closed chain. The term $-iv$ will appear as a boundary condition
for the self energy and will play a crucial role, as demonstrated
below.

In terms of the locator expansion \cite{LGP, PW}
$\tilde\Sigma_n(z)$ can be represented as a sum of paths which
start at site $n$, visit other sites and then return (only once!)
to the starting point $n$. Since the paths consist of steps
connecting nearest neighbors (to the left or to the right), it is
clear that $\tilde\Sigma_n(z)$ can be decomposed into two pieces,
``left" and ``right":
\begin{equation}\label{selfenergy3}
\tilde\Sigma_n(z) = \tilde L_n(z) +  R_n(z)
\end{equation}
The ``left" self-energy, $\tilde L_n(z)$, depends only on the
energies of sites to the left of site $n$, i.e., on $\epsilon_j$'s
with $1\leq j<n$. Similarly, $R_n$ depends only on $\epsilon_j$'s
with $j>n$ and, thus, it is ignorant about the fact that the chain
is coupled to the outside world, via site 1 (that is why the $R_n$
are not tilded). The reason for decomposing $\tilde\Sigma_n(z)$
into $\tilde L_n(z)$ and $R_n$ is that these quantities obey
simple recursion relations, which can be iterated to obtain their
probability distributions:
\begin{equation}\label{recursion1}
\tilde L_j(z)= [z-\epsilon_{j-1}-\tilde L_{j-1}(z)]^{-1}
\end{equation}
and
\begin{equation}\label{recursion2}
R_j(z)= [z-\epsilon_{j+1}-R_{j+1}(z)]^{-1}
\end{equation}
Relation (\ref{recursion1}) has to be iterated starting with
$j=2$, with the ``initial condition" $\tilde L_1=-iv$. Relation
(\ref{recursion2}), for the semi-infinite chain considered in this
paper, leads to a stationary, $n$-independent distribution for the
variable $R_n$. We will not need this distribution in the
forthcoming calculation, restricted to the case of strong
disorder.

\section{Strong Disorder}
Strong disorder means that a typical value of $\epsilon_n$ is much
larger than $t=1$, i.e., the distribution of site energies,
$q(\epsilon)$, is very broad. In this case the above recursion
relations simplify considerably. The real part of the self-energy
can be neglected in comparison with $\epsilon_n$, so that at any
site $n$ the real part of the resolvent is $Re\tilde{\bf
G}_{nn}(z)\approx (x-\epsilon_n)^{-1}$, which is the same as for
the closed chain. Thus, the real parts of the two resolvents in
(\ref{Gausslaw1}) cancel out and one obtains
\begin{equation}\label{Gausslaw2}
\langle \rho(x,y) \rangle = -{1\over 2\pi}\partial_y\langle {\rm
Tr}Im (\tilde{\bf G}-{\bf G}) \rangle\,.
\end{equation}
Furthermore,  $Im\tilde{\bf G}_{nn}$ (see (\ref {selfenergy2})) is
dominated by $Im\tilde L_n(z)$ and can be approximated as
\begin{equation}\label{G1}
Im\tilde{\bf G}_{nn}={I_n-y\over (x-\epsilon_n)^2 + (I_n
-y)^2}\approx \pi \delta(x-\epsilon_n)sign(I_n-y)
\end{equation}
where, in order to avoid cluttering the notations, we denote
$Im\tilde L_n(z)\equiv I_n$. Since $\epsilon_n$ and $I_n$ are
independent random variables, the averaging of (\ref{G1}) is done
in two steps: first over $\epsilon_n$, which gives the strong
disorder limit of the density of states, $q(x)$, and then over
$I_n$, with the yet unknown distribution $\tilde{P}_n(I)$. The
latter is governed by the recursion relation (\ref{recursion1}),
which in the strong disorder limit simplifies to
\begin{equation}\label{recursion3}
I_j={I_{j-1}\over (x-\epsilon_{j-1})^2}.
\end{equation}
This equation, as well as the previous one, applies for $n\geq2$.
Site $n=1$ provides the initial condition $I_1=-v$. The recursion
relation (\ref{recursion3}) ``propagates" the information from the
open end of the chain (site 1) to distant sites. Site $1$ is
special and should be excluded from the trace in
(\ref{Gausslaw2}). This is because $I_1$ has a fixed value,
different from $y$. Using the electrostatic analogy, one can say
that a charge, located away from the point ($x,y$) cannot produce
a singularity at this point . Actually, (\ref{recursion3}) is
valid only up to a point when $I$ becomes of order $y$ ( recall
that we are interested in $y$ fixed but arbitrarily small, i.e.,
in the ``small-$y$" asymptotic behavior of $\langle \rho(x,y)
\rangle$). When such a small value of $I$ is reached, the initial
condition, i.e., the value of $v$, is forgotten, and there is no
contribution to $\langle \rho(x,y) \rangle$) (more precisely, both
$Im \tilde{\bf G}$ and $Im {\bf G}$ become very small and,
moreover, cancel each other). In summary,  averaging (\ref{G1})
with the distribution $\tilde{P}_n(I)$ and taking the derivative
with respect to $y$, one obtains from (\ref{Gausslaw2}) (with the
${\bf G}$-term neglected):
\begin{equation}\label{Gausslaw3}
\langle \rho(x,y) \rangle=q(x)\sum_{n=2}^{\infty}
{\tilde{P}_n(y)}.
\end{equation}

Eq. (\ref{Gausslaw3}), supplemented by the recursion relation
(\ref{recursion3}), enables us to obtain the asymptotic
($1/y$)-behavior, as we now explain. Note that the random
variables $I_n$ are negative, so that resonances are located in
the lower half-plane of the complex variable $z$. It is more
convenient to work with the absolute value, $|I_n|$, and to define
$y$ as positive, i.e., $\langle \rho(x,y) \rangle$ is the average
density of resonances with width $y$ (at energy $x$). Defining,
instead of $|I_n|$, a new variable, $t_n=ln|I_n|$, and designating
its distribution as $W_n(t)$, we have, instead of
(\ref{Gausslaw3}),
\begin{equation}\label{Gausslaw4}
\langle \rho(0,y) \rangle=q(0){1\over y}\sum_{n=2}^{\infty}
{W_n(t)}, \qquad t=lny,
\end{equation}
where we set $x=0$ (middle of the band). It follows from
(\ref{recursion3}) and the initial condition $|I_1|=v=1$ that
\begin{equation}\label{t}
t_n=-2\sum_{j=1}^{n-1}{ln|\epsilon_j|},
\end{equation}
so that
\begin{equation}\label{W(t)}
W_n(t)=<\int\limits_{c-i\infty}^{c+i\infty} {dp\over 2\pi i}
e^{p(t-t_n)}>=\int\limits_{c-i\infty}^{c+i\infty} {dp\over 2\pi
i}e^{pt}\nu^{j-1}(p),
\end{equation}
where $c$ is some real number and
\begin{equation}\label{nu}
\nu(p)=\int \ d\epsilon q(\epsilon)e^{2pln|\epsilon|}.
\end{equation}
Summing the geometric series in (\ref{Gausslaw4}) we obtain:
\begin{equation}\label{Gausslaw5}
\langle \rho(0,y) \rangle=q(0){1\over
y}\int\limits_{c-i\infty}^{c+i\infty} {dp\over 2\pi
i}e^{pt}{\nu(p)\over 1-\nu(p)},\qquad t=lny.
\end{equation}
This equation makes sense if $|\nu(p)|<1$. A sufficient condition
for this inequality is that $c$ be negative and small,
so that we have to choose the contour somewhat to the left of the
imaginary axis in the complex $p$ plane. In order to prove the
($1/y$)-behavior, in the small $y$ limit, we have to verify that
the integral  in (\ref {Gausslaw5}) approaches a constant
(non-zero) value for $t\rightarrow-\infty$. It is intuitively
quite obvious that in this limit, and for $|c|\ll1$, the integral
will be dominated by the vicinity of the point $p=0$, near which
$\nu(p)\approx1+2p<ln\mid\epsilon\mid>$. For strong disorder,
$<ln\mid\epsilon\mid>$ is positive and equal to the inverse
localization length $1/\xi$ (in the middle of the band). Replacing
in (\ref {Gausslaw5}) ($1-\nu(p)$) by ($-2p/\xi$), we obtain:
\begin{equation}\label{Gausslaw6}
\langle \rho(0,y) \rangle=q(0){1\over
y}\int\limits_{c-i\infty}^{c+i\infty} {dp\over 2\pi
i}e^{pt}{\nu(p)\over (-2p/\xi)}={q(0)\xi\over 2y}.
\end{equation}
Note that, since $t$ is negative, the integration contour should
be closed in the right half-plane, where $Rep$ is positive, so
that an extra minus sign is acquired. The formal proof of
equivalence between (\ref {Gausslaw5}) and (\ref {Gausslaw6}), in
the $t\rightarrow-\infty$ limit, is achieved by studying the
difference between the two expressions. Let's denote by $J$ and
$J'$ the integrals in (\ref {Gausslaw5}) and (\ref {Gausslaw6}),
respectively. In both integrals $c$ cannot be set equal to zero,
because the integrand would diverge at $p=0$. However, this
divergence is cancelled in $J-J'$, so that for this quantity one
can integrate directly along the imaginary $p$-axis. Moreover,
since $\nu(p)$, for $p$ purely imaginary, is just the
characteristic function of the distribution for the variable
$2ln|\epsilon|$, it must decrease faster than $1/|p|$ for large
values of $p$. Therefore one can use the Riemann-Lebesgue lemma to
prove that $J-J'$ approaches zero in the $t\rightarrow-\infty$
limit. This completes the proof of the main result, (\ref
{Gausslaw6}), of our paper. Equation (\ref {Gausslaw6}), gives the
asymptotically exact expression for the average density of
resonances in a semi-infinite, strongly disordered chain.

We are grateful to J. Avron, K. Slevin, T. Kottos, A. Mirlin and,
in particular, to J. Feinberg and E. Gurevich for illuminating
discussions in the course of this work. Thanks are also due to Y.
Fyodorov and M. Titov for some clarifications concerning Ref.
\cite{Tit}. B.S. acknowledges the hospitality of EPFL where most
of this work was done.

\end{document}